\documentclass[twocolumn]{aastex631}

\usepackage{graphicx}% Include figure files
\usepackage{dcolumn}% Align table columns on decimal point
\usepackage{bm}% bold math
\usepackage{xcolor}

\begin{document}

\title{Spatial intermittency of particle distribution in relativistic plasma turbulence}

\correspondingauthor{Cristian Vega}
\email{csvega@wisc.edu}

\author{Cristian Vega}
\affiliation{Department of Physics, University of Wisconsin at Madison, Madison, Wisconsin 53706, USA}

\author[0000-0001-6252-5169]{Stanislav Boldyrev}
%\author{Stanislav Boldyrev}
\affiliation{Department of Physics, University of Wisconsin at Madison, Madison, Wisconsin 53706, USA}
\affiliation{Center for Space Plasma Physics, Space Science Institute, Boulder, Colorado 80301, USA}

\author[0000-0003-1745-7587]{Vadim Roytershteyn}
\affiliation{Center for Space Plasma Physics, Space Science Institute, Boulder, Colorado 80301, USA}



\begin{abstract}
Relativistic magnetically dominated turbulence is an efficient engine for particle acceleration in a collisionless plasma. Ultrarelativistic particles accelerated by interactions with turbulent fluctuations form non-thermal power-law distribution functions in the momentum (or energy) space, $f(\gamma)d\gamma\propto \gamma^{-\alpha}d\gamma$, where $\gamma$ is the Lorenz factor. We argue that in addition to exhibiting non-Gaussian distributions over energies, particles energized by relativistic turbulence also become highly intermittent in space. Based on particle-in-cell numerical simulations and phenomenological modeling, we propose that the bulk plasma density has log-normal statistics, while the density of the accelerated particles, $n$, has a power-law distribution function, $P(n)dn\propto n^{-\beta}dn$. We argue that the scaling exponents are related as $\beta\approx \alpha+1$, which is broadly consistent with numerical simulations. Non-space-filling, intermittent distributions of plasma density and energy fluctuations may have implications for plasma heating and for radiation produced by relativistic turbulence.
\end{abstract}



\section{Introduction}
Particle acceleration in non-equilibrium plasma systems is a fundamental and not completely understood phenomenon of non-linear plasma dynamics. Non-thermal particles are ubiquitous in natural plasmas; they are responsible for gamma rays and X-rays produced in space and astrophysical systems such as solar flares \cite[e.g.,][]{selkowitz2004}, non-thermal radio filaments in the central region of our Galaxy \cite[e.g.,][]{yusef-zadeh2022}, pulsar wind nebulae \cite[e.g.,][]{xu2019,lemoine2016}, and active galactic nuclei jets \cite[e.g.,][]{asano2018}. 

In  weakly collisional plasmas, particles can be accelerated by a variety of mechanisms.   
%Stochastic Fermi accelerated electrons may be responsible for solar flares \cite[e.g.,][]{selkowitz2004}. Non-thermal radio filaments determined in the central region of our Galaxy correspond to magnetic filaments lit up by synchrotron radiation from accelerated electron cosmic rays \cite[][]{}.   Non-thermal relativistic electrons may be responsible for synchrotron radiation coming from pulsar wind nebulae in radio and X-ray diapasons \cite[e.g.,][]{xu2019,lemoine2016}.   Non-thermal emissions from blazars (face-on viewed Active Galactic Nuclei jets) are possibly caused by the electrons accelerated in magnetic turbulence produced in jets \cite[e.g.,][]{asano2018}. Ultra high-energy cosmic rays (UHECRs) ($E>3\times10^{18}$~eV) may possibly be produced due to acceleration of protons in AGN jets or in GRB jets~\cite[e.g.,][]{asano2016}. 
Efficient particle acceleration can be provided by collisionless shocks, where converging magnetized plasma flows act as magnetic mirrors accelerating fast particles bouncing between them \cite[e.g.,][]{blandford1987,bykov1996,marcowith2016}, the so-called first-order Fermi acceleration mechanism. Particles can also be stochastically accelerated by reflections from randomly moving magnetic mirrors, in the second-order Fermi acceleration process \cite[e.g.,][]{fermi1949,teller1954,kulsrud1971,selkowitz2004,petrosian2004}. More recently, it has been realized that processes of magnetic reconnection may efficiently accelerate electrons, arguably due to non-ideal electric fields present in the reconnection layers, motional electric fields associated with plasma outflows, and particles being trapped in contracting and merging plasmoid islands formed in current sheets \cite[e.g.,][]{uzdensky2011,drake2013,sironi2014,guo2020,french2022,sironi2022}. In a magnetically dominant plasma (where the magnetic energy exceeds the rest-mass energy of the particles), reconnection is able to produce non-thermal particles with hard ultra-relativistic energy spectra \cite[e.g.,][]{sironi2014,guo2020}. 
%Astrophysical objects such as PWN, jets from AGNs are studied by their radiation properties. Power-law radiation spectra are signatures of the presence of ultra-relativistic electrons with power-law distribution functions.  

%Non-thermal particle acceleration is a central topic in astrophysics. Particles can be accelerated by different mechanisms. Commonly invoked reconnection mechanisms include particle interaction with collisionless shocks, magnetic reconnection sites. 

Recently, attention has been drawn to particle acceleration by turbulence and specifically, magnetically dominated turbulence, which may provide a complementary or alternative mechanism to previously studied acceleration processes \cite[][]{zhdankin2017a,zhdankin2018,zhdankin2018c,zhdankin2019,zhdankin2021,comisso2018,comisso2019,comisso2021,comisso2022,zhdankin2020,nattila2020,nattila2022,trotta2020,vega2022b,lemoine2022}.  Kinetic numerical simulations indicate that turbulence accelerates particles to ultra-relativistic energies, with power-law energy distribution functions. For strong magnetization, the distribution functions of accelerated particles are observed to be relatively universal: they depend only on the relative strengths of the uniform and fluctuating parts of the magnetic field \cite[e.g.,][]{comisso2019,vega2022a}. The mechanism of particle acceleration by relativistic turbulence as well as the properties of such turbulence itself are currently not well understood. It is however known that turbulent flows are spatially and temporarily intermittent \cite[e.g.,][]{zhdankin2012b,zhdankin2014,zhdankin2016,zrake2013}; they spontaneously generate magnetic structures that trap and scatter particles, as well as magnetic discontinuities and current sheets that trigger reconnection events \cite[e.g.,][]{matthaeus_turbulent_1986,loureiro2017,loureiro2018,loureiro2019,mallet2017a,comisso2019,boldyrev_2017,boldyrev_loureiro2018,boldyrev2019,vega2020,dong2022,pezzi2022}. Interactions of particles with turbulent structures may play a role in the acceleration process \cite[e.g.,][]{vega2022a}. The universality of the resulting particle energy distribution function may be related to the universality of underlying turbulence.

In this work, we demonstrate that particle heating and acceleration in relativistic turbulence are associated with strong spatial intermittency of the resulting particle distribution function. We argue that the acceleration process creates two populations of particles, which have essentially different intermittency characteristics. The bulk of the particles, corresponding to the thermal energies $\gamma \lesssim \gamma_{th}$, develops the log-normal distribution of the particle density.\footnote{In relativistic dynamics, the energy of a particle is given by ${\cal E}=\gamma mc^2$. We can therefore characterize the energy distributions of particles by the distributions of their corresponding gamma factors~$\gamma$.} The population of the accelerated non-thermal particles ($\gamma \gg \gamma_{th}$), on the other hand, has a strongly inhomogeneous ``clumpy" spatial distribution, corresponding to a power-law density distribution function. Strong spatial intermittency of accelerated relativistic particles may be relevant for the radiation flares in such astrophysical objects as pulsar wind nebulae, magnetospheres of black-hole accretion disks, magnetospheres of neutron stars, and blazar jets \cite[e.g.,][]{zhdankin2020,nattila2020,groselj2023}. 

In what follows, we first present general physical arguments suggesting that the distribution of plasma particles energized by relativistic turbulence, should be spatially intermittent. We then compare the results with 2D and 3D particle-in-cell (PIC) numerical simulations of magnetically dominated turbulence in a pair plasma.

\section{Magnetically dominated turbulence: a phenomenological picture}
\label{pheno}
Consider a setting where turbulence is driven by initially strong large-scale magnetic perturbations. We assume that the initial energies of the perturbations $\delta B_{0}$ and the guide magnetic field $B_0$ are much larger than the initial energy (rest-mass plus kinetic) of the plasma particles. We also assume that the guide field is not too strong as compared to the fluctuations, $B_0\lesssim \delta B_0$. As the magnetic perturbations relax, their energy is converted into relativistic large-scale hydrodynamic plasma flows, whose Lorentz factors $\tilde \gamma$ may locally significantly exceed unity. Such ultra-relativistic plasma flows are however highly compressible, and as was pointed out in \cite{vega2022a,vega2022b}, their velocities rapidly (on a few large-scale crossing times) become sub-relativistic.  Indeed, the speed of sound in a relativistic plasma does not exceed $c/\sqrt{3}$ (assuming adiabatic large-scale motion), so initial ultrarelativistic and, therefore, supersonic flows rapidly relax to subsonic velocities $\lesssim c/\sqrt{3}$, while simultaneously, the plasma is heated to ultrarelativistic temperatures. This phenomenological picture is consistent with available numerical observations \cite[e.g.,][]{zhdankin2018c,vega2022a}, and with the numerical results presented below.

The resulting relativistic turbulence is therefore inherently compressible. Compressible turbulence is associated with spatially intermittent density fluctuations.   In a hydrodynamic picture, the probability distribution function of density fluctuations acquires the log-normal statistics.  This may be illustrated in a simple hydrodynamic model of Gaussian random advection. Assume, somewhat idealistically, that the velocity field is non-relativistic, random, Gaussian, and independent of the density fluctuations. For simplicity, we also assume that this field is short-time correlated, and has a zero mean, $\langle {\bm v}({\bf x}, t) \rangle=0$. Its statistics are therefore fully described by the covariance
\begin{eqnarray}
\langle v^i({\bf x}, t)v^j({\bf x}', t')\rangle=2\kappa^{ij}({\bf x}-{\bf x}')\delta(t-t'),  
\label{eqN}
\end{eqnarray}
where the tensor $\kappa^{ij}({\bf x}-{\bf x}')$ describes spatial correlation of the velocity fluctuations, and $\delta(t-t')$ is the Dirac delta function. Starting from the continuity equation for the density field, 
\begin{eqnarray}
\frac{\partial n}{\partial t}+{\bm \nabla}\cdot (n{\bm v})=0,
\end{eqnarray}
which can be viewed as a stochastic Langevin equation with the random noise ${\bm v}({\bf x}, t)$, one can derive the corresponding Fokker-Planck equation for the probability density function of the density field,
\begin{eqnarray}
\label{probaN}
\frac{\partial P(n,t)}{\partial t}=D \frac{\partial}{\partial n} n \frac{\partial}{\partial n}\left(n P \right).    
\end{eqnarray}
Here, we introduced the diffusion coefficient defined as
\begin{eqnarray}
D=\nabla_i\nabla^\prime_j\kappa^{ij}({\bf x}-{\bf x}') \vert_{{\bf x}={\bf x}'},
\end{eqnarray}
and we sum over repeated indexes. The derivation of the Fokker-Planck equation is standard and can be found in e.g., \cite[][Ch. 34]{zinn-justin2021}, and the Stratonovich convention is used. As can be directly verified, the solution to the diffusion equation (\ref{probaN}) is the log-normal distribution. Numerical simulations of non-relativistic compressible turbulence indeed produce log-normal density distributions \cite[e.g.,][]{kritsuk2017}. Below we will demonstrate that such distributions are also observed in our numerical simulations of relativistic turbulence.

A small population of plasma particles, however, get accelerated by turbulent electric fields in a run-away fashion, in that their energy distribution significantly deviates from the equilibrium Maxwell-J\"uttner distribution and develops power-law tails \cite[e.g.,][]{zhdankin2018,comisso2018}. We argue that this population of accelerated, ultra-relativistic particles does not, in general, move with the bulk velocity of the plasma but rather tends to cluster in space. We observe that the statistics of such particles are strongly intermittent, i.e., described by power-law probability density functions, not only in the momentum but also in configuration space.

Numerical simulations indicate that particle acceleration involves two stages \cite[][]{comisso2019,zhdankin2020}. At the first stage, particles are accelerated by reconnection events, which are obviously intermittent in space. At the second stage, particles are accelerated by interaction with random turbulent fluctuations. One can argue that acceleration by turbulence may also lead to spatially intermittent particle distributions.

\begin{figure}[tbh!]
\includegraphics[width=\columnwidth]{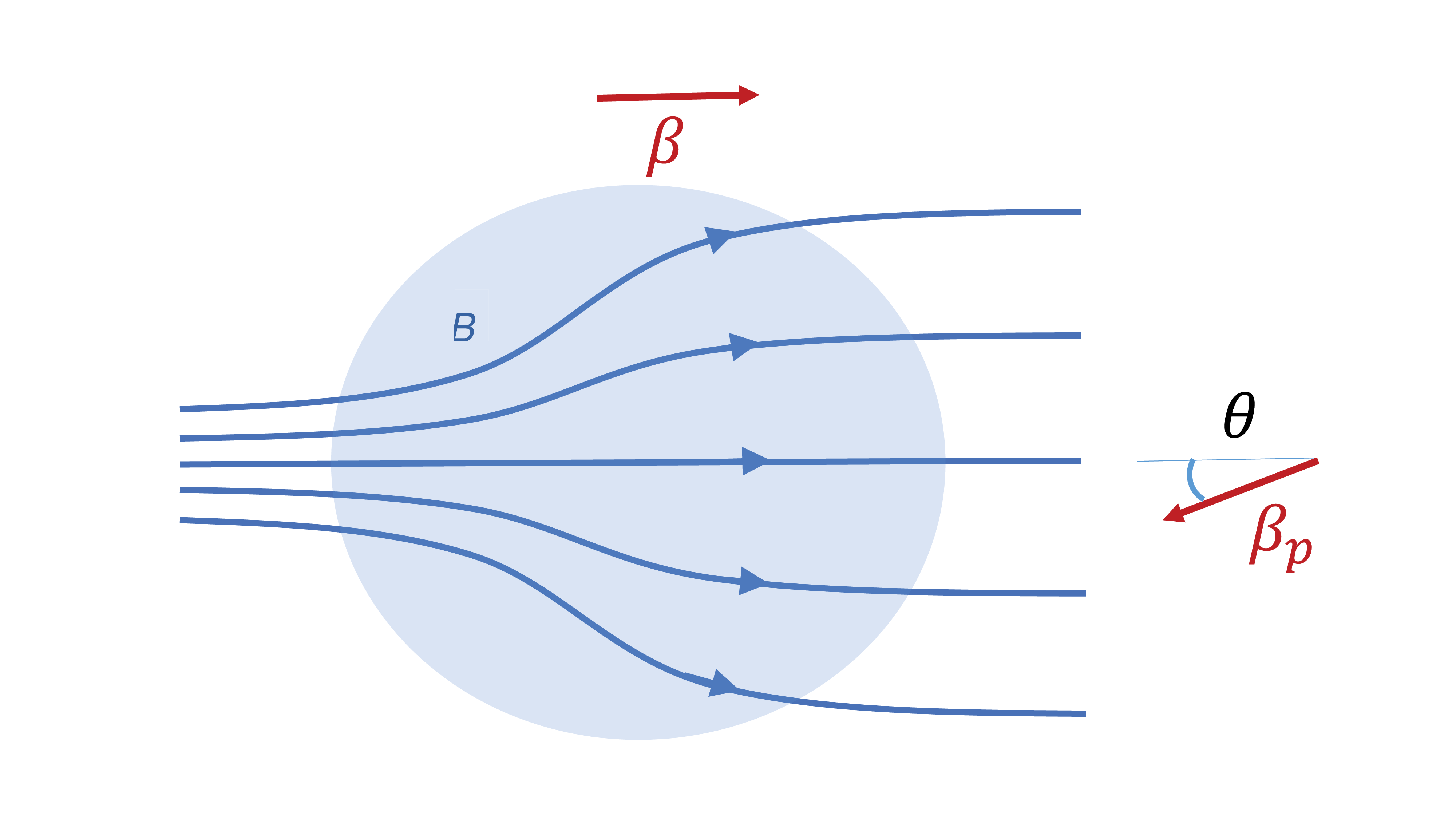}
\caption{Sketh of particle interaction with a strong magnetic mirror moving with a plasma element.}
\label{figure2}
\end{figure}

As an example illustrating that accelerated particles may cluster in space, consider a quasi-one-dimensional model where particles are scattered by a strong magnetic mirror shown in Fig.~\ref{figure2}. Such a mirror may correspond to large-scale magnetic fluctuations in turbulence. Assume that the mirror is moving in the normal direction with the beta factor $\bm{\beta}=\bm{v}/c$, the particles propagate along the axis of the mirror at the pitch angle~$\theta$, and their beta factor is ${\bm \beta}_p=\bm{v}_p/c$. Assume that the particles have similar Lorenz factors $\gamma$ and occupy a small spatial volume $d V$.
In the mirror rest frame, the magnetic field is stationary and the energy of the particles does not change during a reflection. It can be demonstrated by direct calculation that in the laboratory frame the energy of the particles after a reflection is given by
\begin{eqnarray}
&\gamma'=\lambda \gamma,
\label{gamma}
\end{eqnarray}
with the scaling factor 
\begin{eqnarray}
\lambda =\frac{1+\beta^2-2\bm{\beta}\cdot\bm{\beta}_p}{1-\beta^2}. 
\end{eqnarray}

Obviously, the particles get accelerated when ${\bm \beta}_p\cdot {\bm \beta}<0$.
Simultaneously, one can show that the volume of the fluid element gets compressed along the axis of the mirror according to 
\begin{eqnarray}
 d V^\prime =d V/\lambda, 
\label{dV}
\end{eqnarray} implying that the process of acceleration leads to stronger spatial clusterization of particles. Relation (\ref{dV}) holds when particles propagating with the same pitch angle $\theta$, also have the same parallel velocities $\beta_p\cos(\theta)$. This is possible for the ultrarelativistic case  $\beta_p\to 1$. The ultrarelativistic case also ensures that the factor $\lambda$ depends on the first order of $\beta$, which makes the acceleration most efficient.

Similarly, one can address the ``geometric'' dispersion, that is, dispersion due to different pitch angles of particle velocities. One can show that these pitch angles become progressively closer to each other (and to the axis of the mirror) after each reflection with $\lambda>1$. If a beam of ultrarelativistic particles is moving within a small collimation angle $d \theta$ around the angle ${\theta}$, then after the reflection, the new collimation angle satisfies:
\begin{eqnarray}
d \theta'=d \theta/\lambda.
\end{eqnarray}
Reflections where particles get accelerated ($\lambda >1$) lead to stronger angular collimation of the particle beam. 

This is consistent with our picture of the intermittent spatial distribution of accelerated particles. Our consideration may also be relevant to the two interesting observations made in recent numerical simulations of turbulent acceleration. First is that it is necessary to initially accelerate (possibly, by a mechanism different from the second-order Fermi acceleration) the particles to ultrarelativistic energies before their energy distribution function starts to form a universal power-law  tail \cite[][]{comisso2019}. The second observation is that radiation of accelerated particles has the character of ``flares,"  which are intermittent in space and/or direction~\cite[][]{zhdankin2020b}.

Numerical simulations and phenomenological models indicate \cite[e.g.,][]{zhdankin2017a,zhdankin2018,zhdankin2018c,zhdankin2019,zhdankin2021,comisso2018,comisso2019,comisso2021,comisso2022,zhdankin2020,nattila2020,nattila2022,trotta2020,vega2022a,vega2022b,lemoine2022} that accelerated particles have power-law energy distribution functions,
\begin{eqnarray}
f(\gamma)d\gamma \propto \gamma^{-\alpha}\,d\gamma.  \label{fgam}  
\end{eqnarray}
Based on the consideration presented above, we may analyze the probability density functions of the density, $P(n)$, for such particles. For that consider a nearly monoenergetic fluid element consisting of a given number of particles.   As these particles get accelerated, according to Eqs.~(\ref{gamma})~and~(\ref{dV}), their density increases proportionally to their energy, $n\propto \gamma$. The volume occupied by such an element, however, decreases inversely proportionally to~$n$. The probability to observe density $n$ at a given point in space (or the fractional volume with the densities in the interval $[n, n+dn ]$) is therefore:
\begin{eqnarray}
P(n)\,dn \propto n^{-\alpha-1}\,dn,\label{Pn}
\end{eqnarray}
which suggests that the spatial distribution of the accelerated particles is also strongly intermittent and described by power-law probability density functions. In the next section, we verify our phenomenological predictions for the particle distribution function $P(n)$ in first-principle particle-in-cell (PIC) numerical simulations. 

\section{Numerical results}

We conduct 2D (or ``2.5D") and 3D simulations of decaying turbulence in a pair plasma with the fully relativistic particle-in-cell code VPIC \cite[][]{bowers2008}. In the 2.5D simulations, the electric and magnetic fields have three vector components but depend only on two spatial coordinates $x$ and $y$. Similarly, the particle distribution functions depend on three velocity components. We define the two plasma magnetization parameters in terms of the imposed uniform background magnetic field, ${\bf B_0}=B_0{\hat z}$, and the root-mean-square value of initial magnetic fluctuations, $\delta B_{0}=\langle\delta B^2({\bf x}, t=0)\rangle^{1/2}$, as 
\begin{eqnarray}
\sigma_0=\frac{B_0^2}{4\pi n_0w_0mc^2} 
\end{eqnarray}
and
\begin{eqnarray}
{\tilde \sigma}_0=\frac{(\delta B_0)^2}{4\pi n_0w_0mc^2}.
\end{eqnarray}
In these expressions, $n_0$ denotes the initial uniform density of {each} plasma species (electrons and positrons), 
%$w_0mc^2=[K_3(1/\theta_0)/K_2(1/\theta_0)]mc^2$ is the initial enthalpy per particle, 
$w_0=K_3(1/\theta_0)/K_2(1/\theta_0)$ is the normalized initial enthalpy per particle, where 
$K_\nu(z)$ is the modified Bessel function of the second kind, and $\theta_0=k_BT_0/mc^2$ is the normalized initial temperature (for ultrarelativistic temperatures, $\theta_0\gg 1$, one has $w_0\approx 4\theta_0$, while for non-relativistic plasma, $\theta_0\ll 1$,  $w_0\approx 1$). 
%\footnote{It is easy to see that for ultrarelativistic temperatures, $\theta_0\gg 1$, one gets $w_0\approx 4\theta_0$, while for a nonrelativistic plasma, $\theta_0\ll 1$, one gets $w_0\approx 1$.}. 
The initial distribution of plasma particles is an isotropic Maxwell-J\"uttner distribution with the temperature $\theta_0=0.3$. The parameters of the runs are summarized in Table~{\ref{table}}.

\begin{table}[h!]
%\vskip5mm
\hskip-2.0cm
\centering
\begin{tabular}{c c c c c c c} 
\hline
Run & Dim. & $\sigma_0$ & ${{\tilde \sigma}_0 }$ & $\left({B_0}/{\delta B_0}\right)^2$ &  $\omega_{pe}\delta t$ \\
\hline
I & 2D & 2.5 & 40 & 1/16 & 0.04 \\ 
II & 2D & 40 & 40 & 1 & 0.02 \\ 
III & 3D & 40 & 40 & 1 & 0.03 \\
IV & 3D & 80 & 80 & 1 & 0.03 \\
\hline
\end{tabular}
\caption{Parameters of the runs.}
\label{table}
\end{table}

For the 2.5D runs, the simulation domain is a double periodic square with sides $L_x=L_y\approx 2010 \,d_e$, where $d_e$ is the non-relativistic electron inertial scale, with resolution $N_x=N_y=16640$. For the 3D simulations, the domain is a triple periodic cube with sides $L_x=L_y=L_z\approx 1005 \,d_e$, with resolution $N_x=N_y=N_z=2048$. The time steps are normalized to the {inverse non-relativistic} electron plasma frequency, $1/\omega_{{pe}}$. With such parameters, the inertial scale is resolved in both 2.5D and 3D simulations. The initial gyroradius in resolved  in 2.5D but unresolved in 3D runs. Both 2.5D simulations have 100 particles per cell per species and both 3D simulations have 16 particles per cell per species. For the data analysis of the 3D simulations, reduced resolution data was used, with each of the lower resolution cells being eight of the full-resolution cells, leaving the average number of particles per cell at  $8\times16=128$, which is approximately the same as in the 2.5D cases. Apart from that, no additional processing of the data such as smoothing or filtering, has been used.

The turbulence is seeded by randomly phased magnetic perturbations of the type
\begin{eqnarray}
\delta\mathbf{B}(\mathbf{x})=\sum_{\mathbf{k}}\delta B_\mathbf{k}\hat{\xi}_\mathbf{k}\cos(\mathbf{k}\cdot\mathbf{x}+\phi_\mathbf{k}),
\end{eqnarray}
where the unit polarization vectors are normal to the background field, $\hat{\xi}_\mathbf{k}=\mathbf{k}\times\mathbf{B}_0/|\mathbf{k}\times\mathbf{B}_0|$. The wave vectors of the modes are given by $\mathbf{k}=\{2\pi n_x/L_y,{2}\pi n_y/L_y\}$, where $n_x,n_y=1,...,8$ for the 2.5D runs, and by $\mathbf{k}=\{{2}\pi n_x/L_y,2\pi n_y/L_y,2\pi n_z/L_z\}$, where $n_x,n_y=1,...,4$, $n_z=1,2$ for the 3D runs. All modes  have the same amplitudes $\delta B_{\mathbf{k}}$. In the figures below, time is normalized to the large-scale crossing time $l/c$, where $c$ is the speed of light and the outer scale of turbulence is evaluated as $l={2}\pi/k_{x,y}\,(n_{x,y,\text{max}})=L_{x,y}/n_{x,y,\text{max}}$, with $n_{x,y,\text{max}}=8$ in the 2.5D runs and with $n_{x,y,\text{max}}=4$ in the 3D runs.

\begin{figure}[tb!]
\includegraphics[width=\columnwidth,height=0.7\columnwidth]{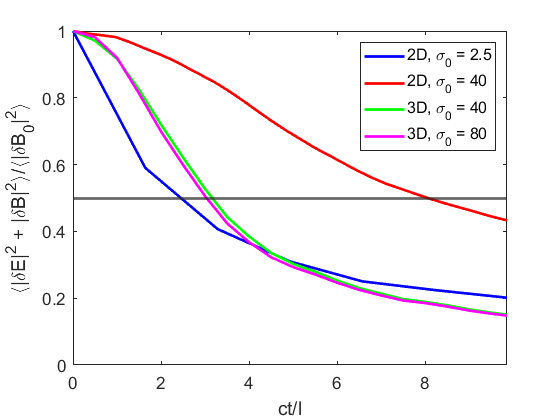}
\includegraphics[width=\columnwidth,height=0.7\columnwidth]{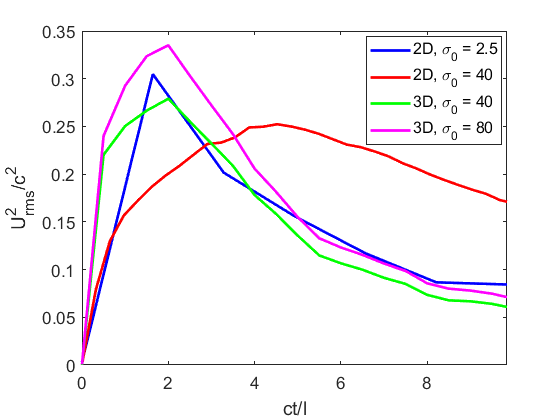}
\caption{Top: Time evolution of energy in electromagnetic fluctuations normalized to its initial value. The horizontal line is at half the initial energy. Bottom: Time evolution of root-mean-square velocity of bulk plasma fluctuations.}
\label{EM_t}
\end{figure}

Figure~\ref{EM_t} shows the time evolution of energy of electromagnetic fluctuations as well as of the bulk velocity of plasma fluctuation in our runs. As turbulence evolves, the initial energy of magnetic fluctuations is transferred to the plasma particles. Since in our numerical setup the initial magnetic energy significantly exceeds the kinetic energy of the particles, by the time the field energy declines by half (we denote this time by $t_{1/2}$), the particles become heated significantly and their kinetic energy becomes comparable to the energy of electromagnetic fluctuations. At approximately the same time, the bulk plasma fluctuations relax to the subsonic velocities $U_{rms} \lesssim c/\sqrt{3}$, in agreement with the phenomenological discussion in section~\ref{pheno}, and the turbulence reaches a quasi-steady state described by universal statistical characteristics such as fluctuations spectra and particle energy distribution functions, in agreement with our previous studies \cite[][]{vega2022b,vega2022a}.

\begin{figure}[tb!]
\includegraphics[width=\columnwidth,height=0.7\columnwidth]{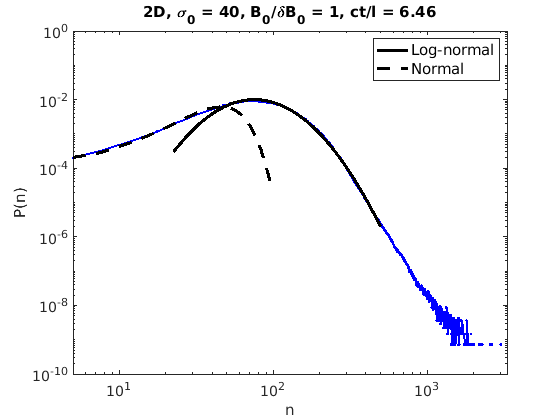}
\includegraphics[width=\columnwidth,height=0.7\columnwidth]{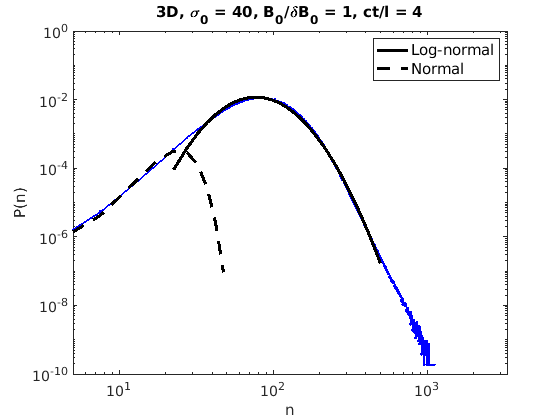}
\caption{Probability density function of particle density $n$ (number of particles per cell) for 2.5D run II (top) and 3D run III (bottom) simulations, both with $\sigma_0=40$. Both curves are well approximated by a log-normal distribution around the peak and a normal distribution at low particle density ($n\ll100$).}
\label{log-normal}
\end{figure}

Figure~\ref{log-normal} shows the distribution of the plasma density obtained in our numerical simulations. 
 
The bulk of the distribution function has intermittent (non-Gaussian) statistics and seems to be well approximated by the log-normal distribution, which is consistent with the prediction of the previous section. At low densities, the numerical curve is, however, better approximated by the normal distribution. This may reflect the fact that the density measurements in particle-in-cell simulations are affected by statistical noise that becomes progressively stronger when particle occupation numbers per cell become progressively smaller.\footnote{We note that VPIC uses first-order particle interpolation.}

\begin{figure*}[h!]
\centering
\includegraphics[width=0.98\columnwidth,height=0.64\columnwidth]{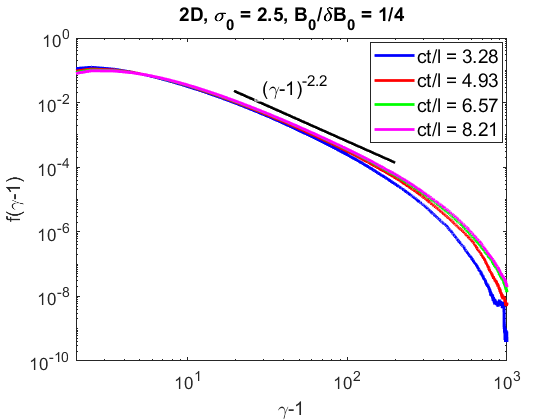}
\includegraphics[width=0.98\columnwidth,height=0.64\columnwidth]{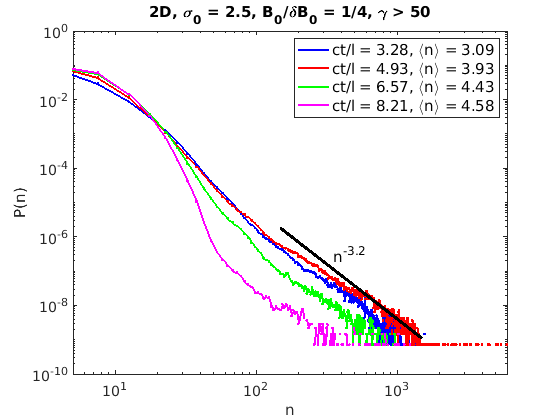}
\includegraphics[width=0.98\columnwidth,height=0.64\columnwidth]{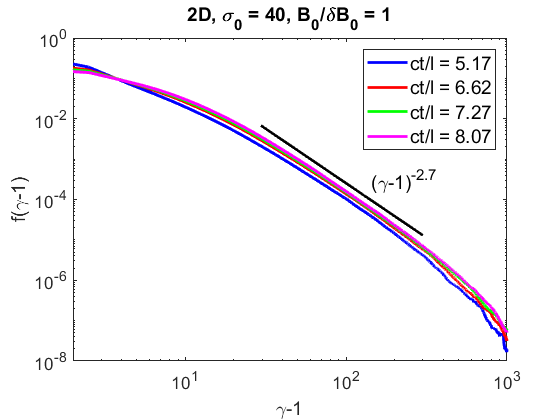}
\includegraphics[width=0.98\columnwidth,height=0.64\columnwidth]{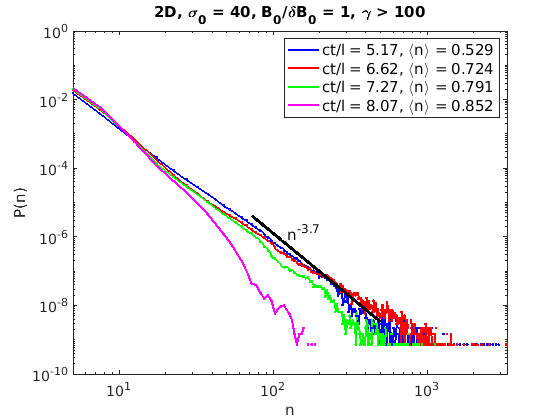}
\includegraphics[width=0.98\columnwidth,height=0.64\columnwidth]{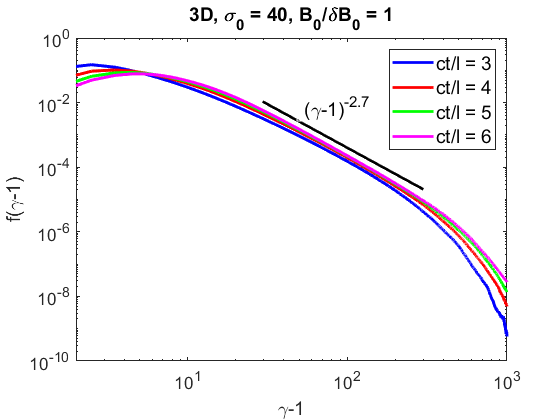}
\includegraphics[width=0.98\columnwidth,height=0.64\columnwidth]{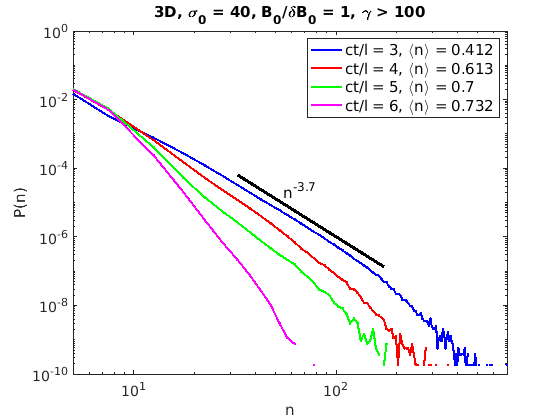}
\includegraphics[width=0.98\columnwidth,height=0.64\columnwidth]{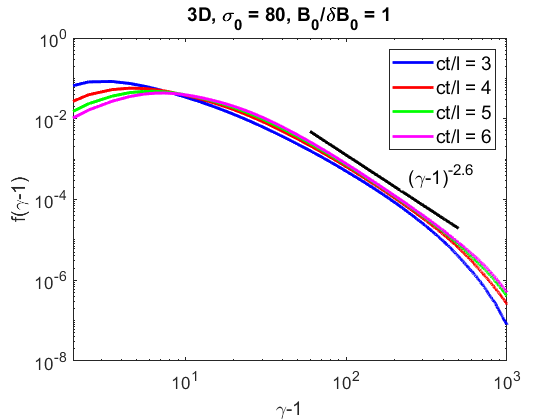}
\includegraphics[width=0.98\columnwidth,height=0.64\columnwidth]{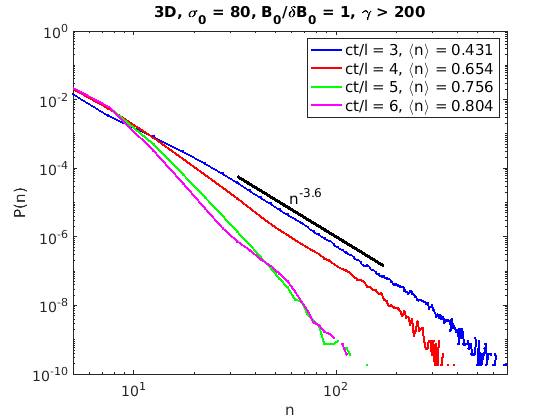}
\caption{Left column: Particle energy pdfs. Quasi-steady states with well-established power-law tails are reached within one or two light-crossing times $l/c$ of the electromagnetic fluctuation energy falling off to half of its initial value. Right column: Particle density pdfs for ultrarelativistic particles. Evidence of strong intermittency (non-Gaussianity) can be seen in the large-density power-law tails. The power slopes $\alpha$ and $\alpha+1$ are given for the reader's orientation.}
\label{fast_part_pdf}
\end{figure*}

\begin{figure*}[]
\centering
\includegraphics[width=\columnwidth,height=0.83\columnwidth]{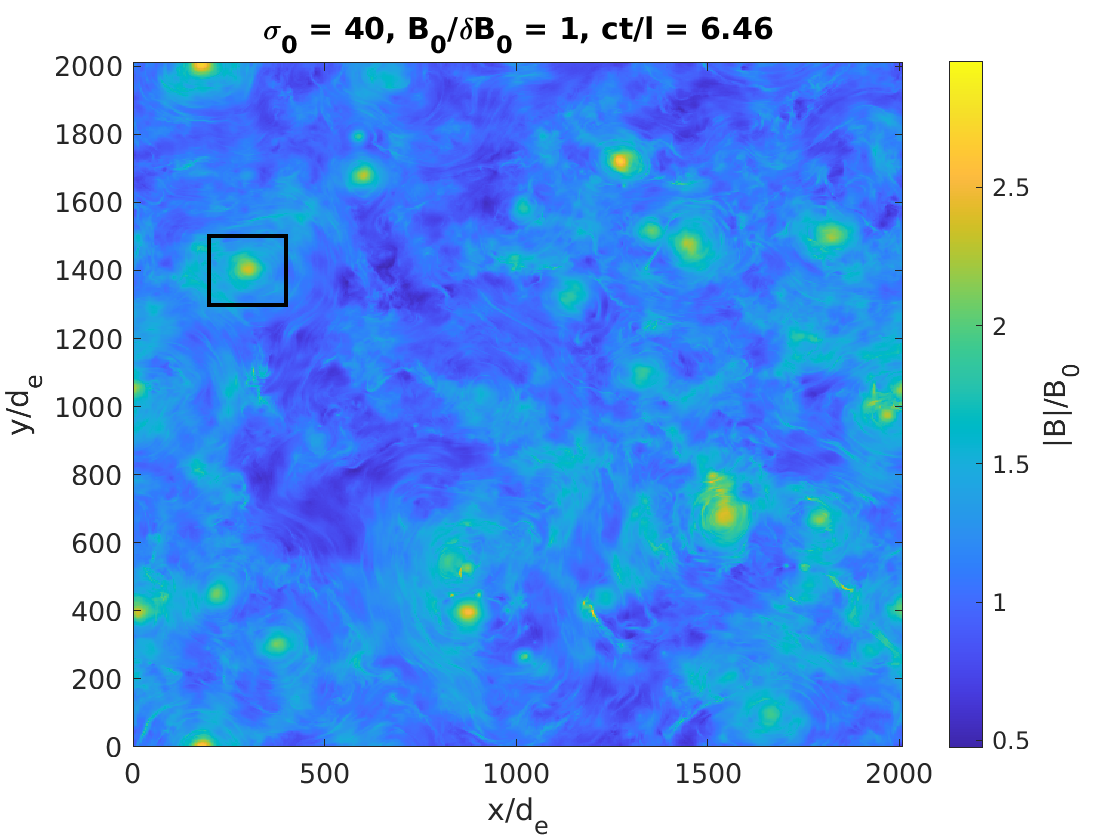}
\includegraphics[width=\columnwidth,height=0.83\columnwidth]{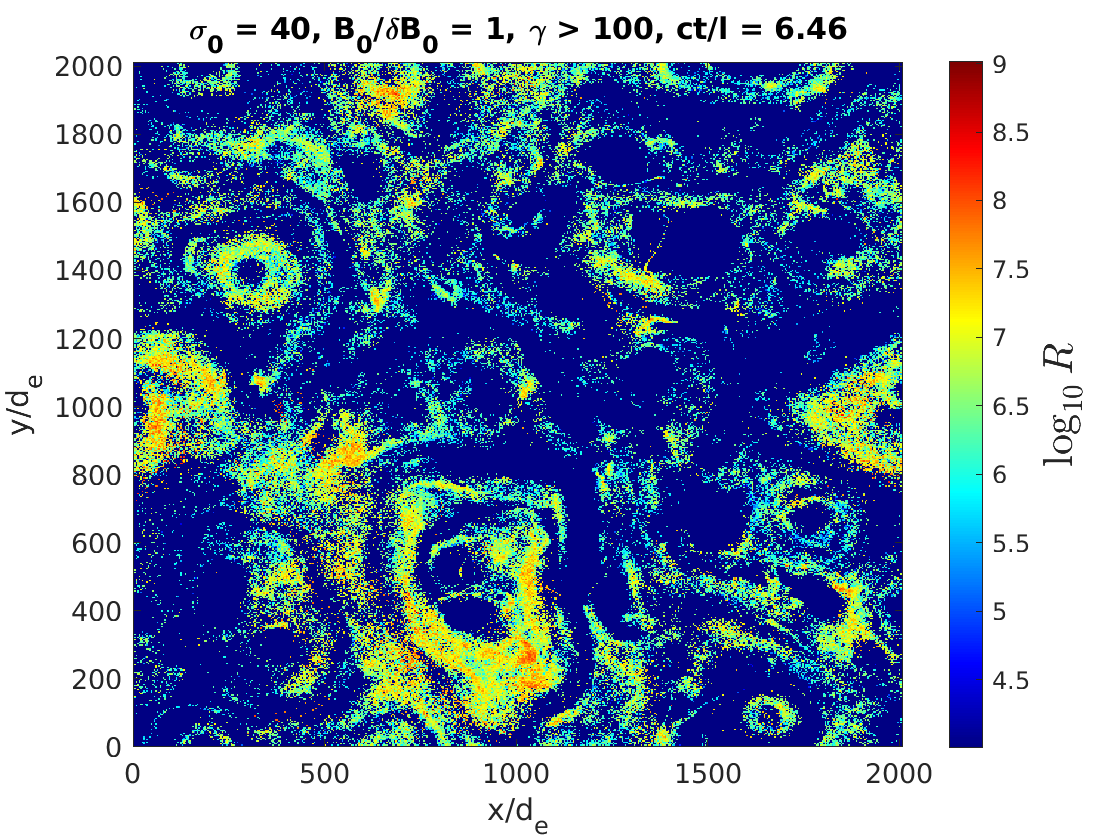}
\includegraphics[width=\columnwidth,height=0.83\columnwidth]{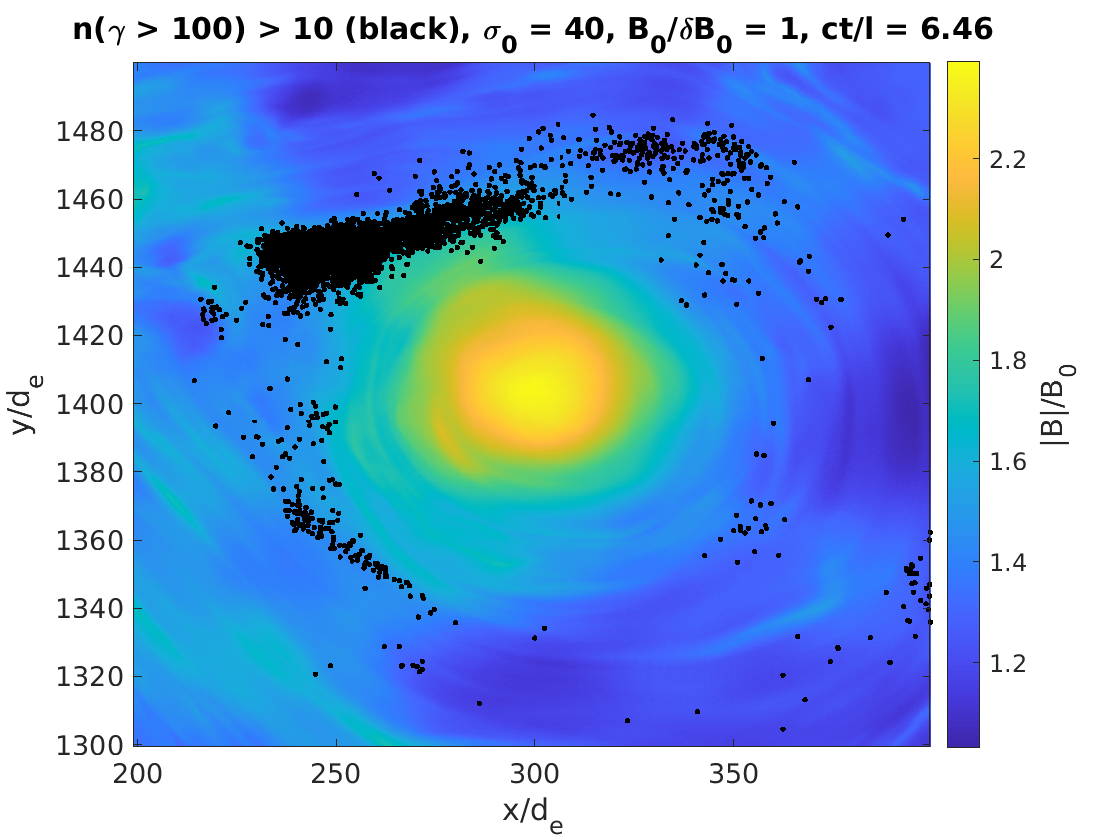}
\includegraphics[width=\columnwidth,height=0.83\columnwidth]{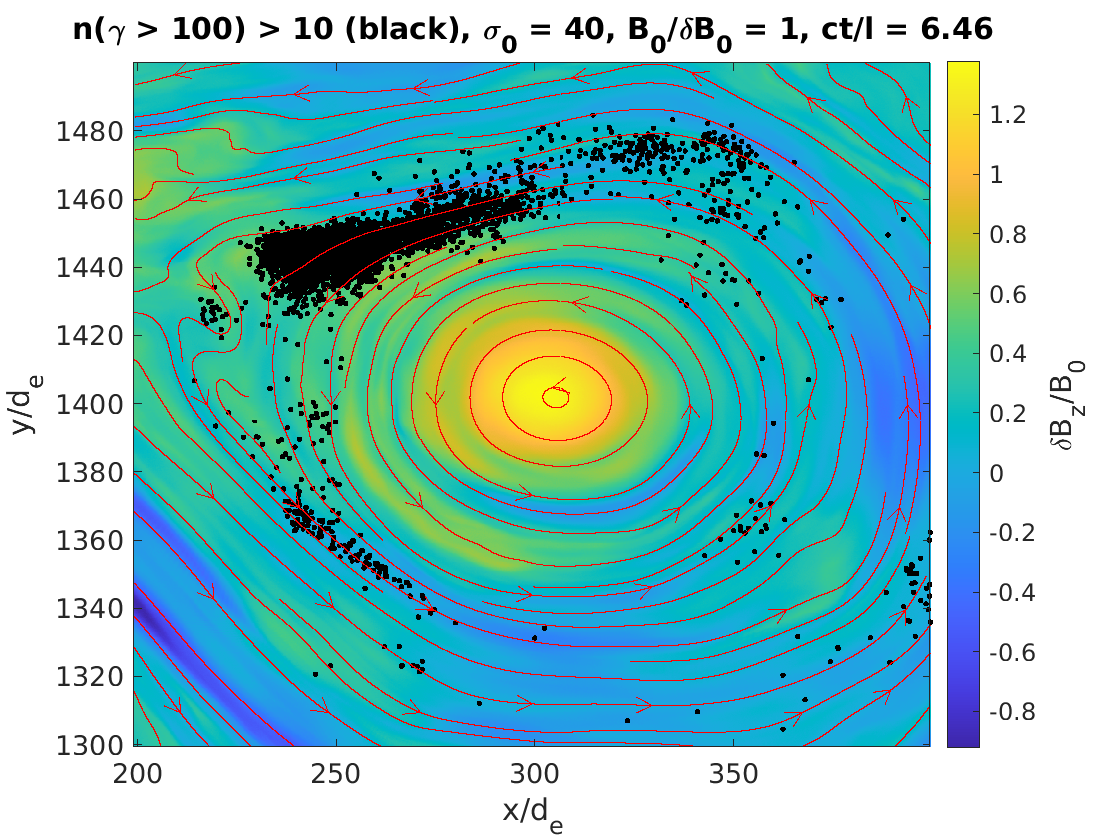}
\caption{Top left: The color map shows the distribution of magnetic-field strength in the simulation domain. Top right: The color map shows  the synchrotron radiation proxy $R=\sum_i\gamma_i^2\beta_i^2\sin^2\theta_i B^2/B_0^2$, where the sum is over the electrons in each cell in Run~II and only the energetic particles with $\gamma>100$ are taken into account. This is a proxy of the synchrotron radiation that would be emitted by electrons if this effect was included in the simulation.   Note the logarithmic scale used in the color code, which implies very strong variations of the intensity of radiation in the simulation domain {(the bottom of the color bar identifies all cells where $R<10^4$)}. Bottom left: An enlarged subdomain indicated by the square in the top left panel. The color map shows the total magnetic field $|\mathbf{B}|$ and the black dots identify cells where there are more than 10 energetic ($\gamma>100$) particles. Bottom right: The color map shows the magnetic fluctuations in the $z$ direction $\delta B_z$. The field lines of the in-plane magnetic field $\delta {\bf B}_\perp$ are shown in read. The clusters of particles seem to concentrate in the vicinity of magnetic structures and align with the magnetic field lines. For reader's orientation, we note that the gyroradius of ultra-relativistic particles scales in the presented simulation as $\rho_e\approx (\gamma/10)d_e$.}
\vskip0.3cm
\label{Rad}
\end{figure*}

The left column of Figure~\ref{fast_part_pdf} shows the time evolution of the particle-energy pdf in each simulation. 
%With the exception of run I in Table~\ref{table} ($B_0/\delta B_0=1/4$), 
A quasi-steady state with a high-energy power-law tail is reached at (or within a dynamical time $l/U_{rms}$ of) the relaxation time $t_{1/2}$,  
suggesting that the energization of ultrarelativistic particles forming the power-law tail happens mostly at $t\lesssim t_{1/2}$, while the bulk of the plasma continues to be heated afterward. 

Also, a comparison with similar runs by \citet[][Figs 9 and 11]{comisso2019} suggests that at the beginning of the initial-magnetic-field relaxation, the particle acceleration is mostly provided by magnetic reconnection events, while by $t\sim t_{1/2}$ the energization process is taken over by interactions with random turbulent fluctuations developed in the system. (This statement should be understood in a statistical sense, meaning that the particle energy distribution function evolves differently in these two regimes. As for individual particles, their acceleration to high energies involves two stages, even in a quasi-steady state developed at~$t_{1/2}$. As demonstrated by \citet{comisso2019}, in the first stage, the so-called initial particle injection, rapid particle acceleration is provided by reconnection events, while in the second stage, a gradual stochastic Fermi-type acceleration is provided by interactions with turbulent fluctuations~\cite[see also][]{zhdankin2020}.) Since in decaying turbulence, the velocity of bulk plasma fluctuations keeps declining at later times, the turbulent acceleration becomes progressively less efficient beyond~$t_{1/2}$ as well. We may therefore conclude  that the most efficient non-thermal particle acceleration by turbulent fluctuations is achieved on a time scale of~$t_{1/2}$.

The right column of Figure~\ref{fast_part_pdf} shows the time evolution of the density pdf calculated only for the energetic particles in each simulation.\footnote{In these measurements, the density is not produced by the code as in Figure~\ref{log-normal}, but rather directly calculated by counting the number of energetic particles in each cell. As the particle population numbers are large, the statistical noise does not noticeably affect the measurements. We verified this by constructing the same density pdfs by using larger cells, where the noise is smaller. We did not observe a significant difference with the presented results.}  We see clear evidence of strong intermittency of density fluctuations associated with the accelerated particles, indicated by power-law tails of the corresponding density distributions. As with Figure~\ref{log-normal}, our 2.5D and 3D simulations (runs II and III) produce qualitatively similar results. Comparison of 3D cases III and IV suggests that the power-law tails become better defined at larger magnetizations~$\tilde{\sigma}_0$. Around the turbulence relaxation times $t_{1/2}$, the observed power laws are broadly consistent with Equation~\ref{Pn}. At $t> t_{1/2}$, that is, after the phase of active particle acceleration, the density pdfs steepen and may be expected to eventually evolve toward the distribution functions of the bulk plasma particles.

The non-Gaussian log-normal or power-law distributions of the density fluctuations are related to the spatially intermittent, non-space-filing structures formed by turbulence. Such structures are known to affect energy dissipation and plasma heating. As an illustrative example, we consider the intermittency of synchrotron radiation that may be generated in relativistic turbulence. We assume that the medium is optically thin and that the synchrotron cooling time significantly exceeds the particle acceleration time. The average power of synchrotron radiation emitted by an ultra-relativistic electron propagating at a pitch angle $\theta$ is given by
\begin{eqnarray}
P=2c \sigma_T\gamma^2\frac{B^2}{8\pi}\beta^2\sin^2\theta,
\end{eqnarray}
where 
\begin{eqnarray}
\sigma_T=\frac{8\pi}{3}\left(\frac{e^2}{m_ec^2} \right)^2 
%\sigma_T=\frac{8\pi}{3}\left(\frac{e^2}{m_ec^2} \right)^2 \approx 6.65\times 10^{-25}\mbox{cm}^2
\end{eqnarray}
is the Thomson cross-section of the electron. Our numerical code does not include radiation. We however note that the radiation power is proportional to $\gamma^2 B^2\beta^2\sin^2\theta$, so we may study the proxy to the density of radiative losses by analyzing the distribution of the local dimensionless quantity $R=\sum_V \gamma_i^2(B/B_0)^2\beta_i^2\sin^2\theta_i$, where we sum over all the electrons contained in a given small volume $V$. The top panels of Figure~\ref{Rad} illustrate the spatial distribution of the magnetic field and the radiation proxy $R$ in Run~II. This Figure suggests that relativistic turbulence generates very strong, spatially intermittent variations of the radiation power density in the simulation domain. For instance, 50\% of the energetic radiation coming from the particles with $\gamma >100$ (the top right panel of Figure~\ref{Rad}), is generated in just {3.3}\% of the volume of the domain, while 80\% of the radiation comes from about 10\% of the volume.

The bottom panels of Figure~\ref{Rad} show a zoomed-in region of the simulation domain, which illustrates the clusterization of energetic particles with $\gamma>100$. Fast particles seem to be concentrated in the vicinity of a strong magnetic structure, and the shapes of their clusters follow the morphology of the magnetic field lines. This may also be consistent with the expectation that the interaction of particles with magnetic structures (or trapping of particles inside such structures) plays an important role in their acceleration \cite[e.g.,][]{vega2022a}.

 \section{Conclusions} Relativistic plasma turbulence is known to be efficient in non-thermal particle acceleration, resulting in power-law  energy distribution functions of highly energetic ultrarelativistic particles $f(\gamma)d\gamma \propto\gamma^{-\alpha}d\gamma$. In this work, we proposed based on particle-in-cell numerical simulations and analytical modeling that magnetically dominated relativistic turbulence is essentially strongly compressible and it naturally generates spatially intermittent distributions of plasma particles. The bulk of the plasma particles, with energies comparable to the average energy (we call it the ``thermal energy''  $\gamma_{th}$), have an essentially non-Gaussian, log-normal density distribution. The ``run-away'' fraction of particles that are accelerated to much higher energies $\gamma\gg \gamma_{th}$, exhibit even more intermittent statistics, with the power-law distribution functions of their number density $P(n)dn\propto n^{-\beta}dn$. 
Based on numerical observations and phenomenological modeling, we argue that the scaling exponents are related approximately as $\beta=\alpha+1$.  %and observe a broad consistency of this relation with the numerical results. 
 
 The strongly non-Gaussian statistics of particle distribution are related to the formation of structures or density ``clumps'' in the simulation domain. Such strong spatial intermittency may have important implications for energy dissipation in relativistic turbulence. As an example, we have considered the distribution of energetic synchrotron radiation that can be produced by particles with $\gamma\gg\gamma_{th}$ in such turbulence. We observed that the contrast of the radiation intensity spans many orders of magnitude over a simulation domain, with the majority of the radiated energy originating in a small fraction of the plasma volume. 

%\smallskip
\begin{acknowledgments}
The work of CV and SB was partly supported by NSF Grant PHY-2010098 and by the Wisconsin Plasma Physics Laboratory (US Department of Energy Grant DE-SC0018266). VR was partially supported by NASA grant 80NSSC21K1692. Computational resources were provided by the Texas Advanced Computing  Center at the University of Texas at Austin (ACCESS Allocation No. TG-ATM180015) and by the NASA High-End Computing (HEC) Program through the NASA Advanced Supercomputing (NAS) Division at Ames Research Center.
\end{acknowledgments}


% \bibliography{sample631}{}
% \bibliographystyle{aasjournal}



\bibliography{references}

\begin{thebibliography}{}
\expandafter\ifx\csname natexlab\endcsname\relax\def\natexlab#1{#1}\fi
\providecommand{\url}[1]{\href{#1}{#1}}
\providecommand{\dodoi}[1]{doi:~\href{http://doi.org/#1}{\nolinkurl{#1}}}
\providecommand{\doeprint}[1]{\href{http://ascl.net/#1}{\nolinkurl{http://ascl.net/#1}}}
\providecommand{\doarXiv}[1]{\href{https://arxiv.org/abs/#1}{\nolinkurl{https://arxiv.org/abs/#1}}}

\bibitem[{{Asano} \& {Hayashida}(2018)}]{asano2018}
{Asano}, K., \& {Hayashida}, M. 2018, \apj, 861, 31,
  \dodoi{10.3847/1538-4357/aac82a}

\bibitem[{{Blandford} \& {Eichler}(1987)}]{blandford1987}
{Blandford}, R., \& {Eichler}, D. 1987, \physrep, 154, 1,
  \dodoi{10.1016/0370-1573(87)90134-7}

\bibitem[{{Boldyrev} \& {Loureiro}(2017)}]{boldyrev_2017}
{Boldyrev}, S., \& {Loureiro}, N.~F. 2017, The Astrophysical Journal, 844, 125,
  \dodoi{10.3847/1538-4357/aa7d02}

\bibitem[{{Boldyrev} \& {Loureiro}(2018)}]{boldyrev_loureiro2018}
{Boldyrev}, S., \& {Loureiro}, N.~F. 2018, in Journal of Physics Conference
  Series, Vol. 1100, Journal of Physics Conference Series, 012003,
  \dodoi{10.1088/1742-6596/1100/1/012003}

\bibitem[{{Boldyrev} \& {Loureiro}(2019)}]{boldyrev2019}
---. 2019, Physical Review Research, 1, 012006 (R),
  \dodoi{https://doi.org/10.1103/PhysRevResearch.1.012006}

\bibitem[{{Bowers} {et~al.}(2008){Bowers}, {Albright}, {Yin}, {Bergen}, \&
  {Kwan}}]{bowers2008}
{Bowers}, K.~J., {Albright}, B.~J., {Yin}, L., {Bergen}, B., \& {Kwan},
  T.~J.~T. 2008, Physics of Plasmas, 15, 055703, \dodoi{10.1063/1.2840133}

\bibitem[{{Bresci} {et~al.}(2022){Bresci}, {Lemoine}, {Gremillet}, {Comisso},
  {Sironi}, \& {Demidem}}]{lemoine2022}
{Bresci}, V., {Lemoine}, M., {Gremillet}, L., {et~al.} 2022, \prd, 106, 023028,
  \dodoi{10.1103/PhysRevD.106.023028}

\bibitem[{{Bykov} \& {Meszaros}(1996)}]{bykov1996}
{Bykov}, A.~M., \& {Meszaros}, P. 1996, \apjl, 461, L37, \dodoi{10.1086/309999}

\bibitem[{{Comisso} \& {Sironi}(2018)}]{comisso2018}
{Comisso}, L., \& {Sironi}, L. 2018, \prl, 121, 255101,
  \dodoi{10.1103/PhysRevLett.121.255101}

\bibitem[{{Comisso} \& {Sironi}(2019)}]{comisso2019}
---. 2019, \apj, 886, 122, \dodoi{10.3847/1538-4357/ab4c33}

\bibitem[{{Comisso} \& {Sironi}(2021)}]{comisso2021}
---. 2021, \prl, 127, 255102, \dodoi{10.1103/PhysRevLett.127.255102}

\bibitem[{{Comisso} \& {Sironi}(2022)}]{comisso2022}
---. 2022, \apjl, 936, L27, \dodoi{10.3847/2041-8213/ac8422}

\bibitem[{{Dong} {et~al.}(2022){Dong}, {Wang}, {Huang}, {Comisso}, {Sandstrom},
  \& {Bhattacharjee}}]{dong2022}
{Dong}, C., {Wang}, L., {Huang}, Y.-M., {et~al.} 2022, Science Advances, 8,
  eabn7627, \dodoi{10.1126/sciadv.abn7627}

\bibitem[{{Drake} {et~al.}(2013){Drake}, {Swisdak}, \& {Fermo}}]{drake2013}
{Drake}, J.~F., {Swisdak}, M., \& {Fermo}, R. 2013, \apjl, 763, L5,
  \dodoi{10.1088/2041-8205/763/1/L5}

\bibitem[{{Fermi}(1949)}]{fermi1949}
{Fermi}, E. 1949, Physical Review, 75, 1169, \dodoi{10.1103/PhysRev.75.1169}

\bibitem[{{French} {et~al.}(2022){French}, {Guo}, {Zhang}, \&
  {Uzdensky}}]{french2022}
{French}, O., {Guo}, F., {Zhang}, Q., \& {Uzdensky}, D. 2022, arXiv e-prints,
  arXiv:2210.08358, \dodoi{10.48550/arXiv.2210.08358}

\bibitem[{{Groselj} {et~al.}(2023){Groselj}, {Hakobyan}, {Beloborodov},
  {Sironi}, \& {Philippov}}]{groselj2023}
{Groselj}, D., {Hakobyan}, H., {Beloborodov}, A.~M., {Sironi}, L., \&
  {Philippov}, A. 2023, arXiv e-prints, arXiv:2301.11327,
  \dodoi{10.48550/arXiv.2301.11327}

\bibitem[{{Guo} {et~al.}(2020){Guo}, {Liu}, {Li}, {Li}, {Daughton}, \&
  {Kilian}}]{guo2020}
{Guo}, F., {Liu}, Y.-H., {Li}, X., {et~al.} 2020, Physics of Plasmas, 27,
  080501, \dodoi{10.1063/5.0012094}

\bibitem[{{Kritsuk} {et~al.}(2017){Kritsuk}, {Ustyugov}, \&
  {Norman}}]{kritsuk2017}
{Kritsuk}, A.~G., {Ustyugov}, S.~D., \& {Norman}, M.~L. 2017, New Journal of
  Physics, 19, 065003, \dodoi{10.1088/1367-2630/aa7156}

\bibitem[{{Kulsrud} \& {Ferrari}(1971)}]{kulsrud1971}
{Kulsrud}, R.~M., \& {Ferrari}, A. 1971, \apss, 12, 302,
  \dodoi{10.1007/BF00651420}

\bibitem[{{Lemoine} {et~al.}(2016){Lemoine}, {Ramos}, \&
  {Gremillet}}]{lemoine2016}
{Lemoine}, M., {Ramos}, O., \& {Gremillet}, L. 2016, \apj, 827, 44,
  \dodoi{10.3847/0004-637X/827/1/44}

\bibitem[{{Loureiro} \& {Boldyrev}(2017)}]{loureiro2017}
{Loureiro}, N.~F., \& {Boldyrev}, S. 2017, \prl, 118, 245101,
  \dodoi{10.1103/PhysRevLett.118.245101}

\bibitem[{{Loureiro} \& {Boldyrev}(2018)}]{loureiro2018}
---. 2018, \apjl, 866, L14, \dodoi{10.3847/2041-8213/aae483}

\bibitem[{{Loureiro} \& {Boldyrev}(2020)}]{loureiro2019}
---. 2020, \apj, 890, 55, \dodoi{10.3847/1538-4357/ab6a95}

\bibitem[{{Mallet} {et~al.}(2017){Mallet}, {Schekochihin}, \&
  {Chandran}}]{mallet2017a}
{Mallet}, A., {Schekochihin}, A.~A., \& {Chandran}, B.~D.~G. 2017, Journal of
  Plasma Physics, 83, 905830609, \dodoi{10.1017/S0022377817000812}

\bibitem[{{Marcowith} {et~al.}(2016){Marcowith}, {Bret}, {Bykov}, {Dieckman},
  {O'C Drury}, {Lemb{\`e}ge}, {Lemoine}, {Morlino}, {Murphy}, {Pelletier},
  {Plotnikov}, {Reville}, {Riquelme}, {Sironi}, \& {Stockem
  Novo}}]{marcowith2016}
{Marcowith}, A., {Bret}, A., {Bykov}, A., {et~al.} 2016, Reports on Progress in
  Physics, 79, 046901, \dodoi{10.1088/0034-4885/79/4/046901}

\bibitem[{{Matthaeus} \& {Lamkin}(1986)}]{matthaeus_turbulent_1986}
{Matthaeus}, W.~H., \& {Lamkin}, S.~L. 1986, Physics of Fluids, 29, 2513,
  \dodoi{10.1063/1.866004}

\bibitem[{{N{\"a}ttil{\"a}} \& {Beloborodov}(2021)}]{nattila2020}
{N{\"a}ttil{\"a}}, J., \& {Beloborodov}, A.~M. 2021, \apj, 921, 87,
  \dodoi{10.3847/1538-4357/ac1c76}

\bibitem[{{N{\"a}ttil{\"a}} \& {Beloborodov}(2022)}]{nattila2022}
---. 2022, \prl, 128, 075101, \dodoi{10.1103/PhysRevLett.128.075101}

\bibitem[{{Petrosian} \& {Liu}(2004)}]{petrosian2004}
{Petrosian}, V., \& {Liu}, S. 2004, \apj, 610, 550, \dodoi{10.1086/421486}

\bibitem[{{Pezzi} {et~al.}(2022){Pezzi}, {Blasi}, \& {Matthaeus}}]{pezzi2022}
{Pezzi}, O., {Blasi}, P., \& {Matthaeus}, W.~H. 2022, \apj, 928, 25,
  \dodoi{10.3847/1538-4357/ac5332}

\bibitem[{{Selkowitz} \& {Blackman}(2004)}]{selkowitz2004}
{Selkowitz}, R., \& {Blackman}, E.~G. 2004, \mnras, 354, 870,
  \dodoi{10.1111/j.1365-2966.2004.08252.x}

\bibitem[{{Sironi}(2022)}]{sironi2022}
{Sironi}, L. 2022, \prl, 128, 145102, \dodoi{10.1103/PhysRevLett.128.145102}

\bibitem[{{Sironi} \& {Spitkovsky}(2014)}]{sironi2014}
{Sironi}, L., \& {Spitkovsky}, A. 2014, \apjl, 783, L21,
  \dodoi{10.1088/2041-8205/783/1/L21}

\bibitem[{{Teller}(1954)}]{teller1954}
{Teller}, E. 1954, Reports on Progress in Physics, 17, 154,
  \dodoi{10.1088/0034-4885/17/1/305}

\bibitem[{{Trotta} {et~al.}(2020){Trotta}, {Franci}, {Burgess}, \&
  {Hellinger}}]{trotta2020}
{Trotta}, D., {Franci}, L., {Burgess}, D., \& {Hellinger}, P. 2020, \apj, 894,
  136, \dodoi{10.3847/1538-4357/ab873c}

\bibitem[{{Uzdensky} {et~al.}(2011){Uzdensky}, {Cerutti}, \&
  {Begelman}}]{uzdensky2011}
{Uzdensky}, D.~A., {Cerutti}, B., \& {Begelman}, M.~C. 2011, \apjl, 737, L40,
  \dodoi{10.1088/2041-8205/737/2/L40}

\bibitem[{{Vega} {et~al.}(2022{\natexlab{a}}){Vega}, {Boldyrev}, \&
  {Roytershteyn}}]{vega2022b}
{Vega}, C., {Boldyrev}, S., \& {Roytershteyn}, V. 2022{\natexlab{a}}, \apjl,
  931, L10, \dodoi{10.3847/2041-8213/ac6cde}

\bibitem[{{Vega} {et~al.}(2022{\natexlab{b}}){Vega}, {Boldyrev},
  {Roytershteyn}, \& {Medvedev}}]{vega2022a}
{Vega}, C., {Boldyrev}, S., {Roytershteyn}, V., \& {Medvedev}, M.
  2022{\natexlab{b}}, \apjl, 924, L19, \dodoi{10.3847/2041-8213/ac441e}

\bibitem[{{Vega} {et~al.}(2020){Vega}, {Roytershteyn}, {Delzanno}, \&
  {Boldyrev}}]{vega2020}
{Vega}, C., {Roytershteyn}, V., {Delzanno}, G.~L., \& {Boldyrev}, S. 2020,
  \apjl, 893, L10, \dodoi{10.3847/2041-8213/ab7eba}

\bibitem[{{Wong} {et~al.}(2020){Wong}, {Zhdankin}, {Uzdensky}, {Werner}, \&
  {Begelman}}]{zhdankin2020}
{Wong}, K., {Zhdankin}, V., {Uzdensky}, D.~A., {Werner}, G.~R., \& {Begelman},
  M.~C. 2020, \apjl, 893, L7, \dodoi{10.3847/2041-8213/ab8122}

\bibitem[{{Xu} {et~al.}(2019){Xu}, {Klingler}, {Kargaltsev}, \&
  {Zhang}}]{xu2019}
{Xu}, S., {Klingler}, N., {Kargaltsev}, O., \& {Zhang}, B. 2019, \apj, 872, 10,
  \dodoi{10.3847/1538-4357/aafb2e}

\bibitem[{{Yusef-Zadeh} {et~al.}(2022){Yusef-Zadeh}, {Arendt}, {Wardle},
  {Boldyrev}, {Heywood}, {Cotton}, \& {Camilo}}]{yusef-zadeh2022}
{Yusef-Zadeh}, F., {Arendt}, R.~G., {Wardle}, M., {et~al.} 2022, \mnras, 515,
  3059, \dodoi{10.1093/mnras/stac1696}

\bibitem[{{Zhdankin} {et~al.}(2012){Zhdankin}, {Boldyrev}, {Mason}, \&
  {Perez}}]{zhdankin2012b}
{Zhdankin}, V., {Boldyrev}, S., {Mason}, J., \& {Perez}, J.~C. 2012, \prl, 108,
  175004, \dodoi{10.1103/PhysRevLett.108.175004}

\bibitem[{{Zhdankin} {et~al.}(2014){Zhdankin}, {Boldyrev}, {Perez}, \&
  {Tobias}}]{zhdankin2014}
{Zhdankin}, V., {Boldyrev}, S., {Perez}, J.~C., \& {Tobias}, S.~M. 2014, \apj,
  795, 127, \dodoi{10.1088/0004-637X/795/2/127}

\bibitem[{{Zhdankin} {et~al.}(2016){Zhdankin}, {Boldyrev}, \&
  {Uzdensky}}]{zhdankin2016}
{Zhdankin}, V., {Boldyrev}, S., \& {Uzdensky}, D.~A. 2016, Physics of Plasmas,
  23, 055705, \dodoi{10.1063/1.4944820}

\bibitem[{{Zhdankin} {et~al.}(2021){Zhdankin}, {Uzdensky}, \&
  {Kunz}}]{zhdankin2021}
{Zhdankin}, V., {Uzdensky}, D.~A., \& {Kunz}, M.~W. 2021, \apj, 908, 71,
  \dodoi{10.3847/1538-4357/abcf31}

\bibitem[{{Zhdankin} {et~al.}(2018{\natexlab{a}}){Zhdankin}, {Uzdensky},
  {Werner}, \& {Begelman}}]{zhdankin2018}
{Zhdankin}, V., {Uzdensky}, D.~A., {Werner}, G.~R., \& {Begelman}, M.~C.
  2018{\natexlab{a}}, \apjl, 867, L18, \dodoi{10.3847/2041-8213/aae88c}

\bibitem[{{Zhdankin} {et~al.}(2018{\natexlab{b}}){Zhdankin}, {Uzdensky},
  {Werner}, \& {Begelman}}]{zhdankin2018c}
---. 2018{\natexlab{b}}, \mnras, 474, 2514, \dodoi{10.1093/mnras/stx2883}

\bibitem[{{Zhdankin} {et~al.}(2019){Zhdankin}, {Uzdensky}, {Werner}, \&
  {Begelman}}]{zhdankin2019}
---. 2019, \prl, 122, 055101, \dodoi{10.1103/PhysRevLett.122.055101}

\bibitem[{{Zhdankin} {et~al.}(2020){Zhdankin}, {Uzdensky}, {Werner}, \&
  {Begelman}}]{zhdankin2020b}
---. 2020, \mnras, 493, 603, \dodoi{10.1093/mnras/staa284}

\bibitem[{{Zhdankin} {et~al.}(2017){Zhdankin}, {Werner}, {Uzdensky}, \&
  {Begelman}}]{zhdankin2017a}
{Zhdankin}, V., {Werner}, G.~R., {Uzdensky}, D.~A., \& {Begelman}, M.~C. 2017,
  \prl, 118, 055103, \dodoi{10.1103/PhysRevLett.118.055103}

\bibitem[{{Zinn-Justin}(2021)}]{zinn-justin2021}
{Zinn-Justin}, J. 2021, {Quantum Field Theory and Critical Phenomena: Fifth
  Edition (5th edn)}, \dodoi{10.1093/oso/9780198834625.001.0001}

\bibitem[{{Zrake} \& {MacFadyen}(2013)}]{zrake2013}
{Zrake}, J., \& {MacFadyen}, A.~I. 2013, \apjl, 763, L12,
  \dodoi{10.1088/2041-8205/763/1/L12}

\end{thebibliography}

\end{document}